# THE ITS-90 AFTER DEFINITION OF NEON ISOTOPIC REFERENCE COMPOSITION: EXTENT OF THE ISOTOPIC EFFECT ON PREVIOUS INTER-COMPARISON RESULTS

Franco Pavese[1*], Anna Szmyrka-Grzebyk[1], Peter P.M. Steur[2]
[1] INTiBS, Wroclaw, Poland
[2] INRIM, Torino, Italy

**Abstract**
Starting from the end of the past century, the importance has been recognised of the effect of isotopic composition on some of the temperature fixed points for the most accurate realisations of the ITS-90. In the original definition of the latter, dating back to 1990, only a generic reference was made to 'natural' composition of the substances used for the realisation of the fixed points, except for helium. The definition of a reference isotopic composition for three fixed points, $e$-$H_2$, Ne and $H_2O$, while eliminating the non-uniqueness of the Scale in this respect, induced detectable differences in the present and future realisations of the Scale, at the highest accuracy level, with respect to the previous realisations, when they affected the results of past MRA key comparisons, namely the CCT K1 (and K1.1) and CCT K2 (and K2.1 to K2.5) and the related regional and supplementary ones. The paper provides evidence of the extent of this effect by using the results of the relevant key comparisons for Neon archived in the BIPM KCDB, and of other comparisons existing in the literature (1979-1984, 2007-2012 and 2009-2010 sealed cell comparisons), and discusses the meaning and the outcomes of this evaluation.

**Keywords:** isotopes; isotopic correction; chemical corrections; neon; triple point; intercomparisons; ITS-90

## 1 Introduction

When the current version of the International Temperature Scale, the ITS-90 [ITS-90], was promulgated in 1990, the isotopic effect on the Scale was basically ignored, except for helium. The Scale definition only made generic reference to a 'natural' composition of the substances used for the realisation of the fixed points.
Since the end of the past century, also thanks to the decreased uncertainty of the best realisations of the ITS-90, the effect of the natural variability of the isotopic composition of some of these substances was recognised as an appreciable contribution to the total uncertainty budget of the realisations of those fixed points, in some cases being the largest single contribution. [Pavese 2005]
Studies were undertaken, initially on $e$-$H_2$ (HD in $H_2$) [Pavese, Tew 2000; Fellmuth *et al.* 2005] and $H_2O$ ($D_2$ and $^{18}O_2$) [Nicholas *et al.* 1996; White et al. 2003], and later also on Ne [Pavese *et al.* 2005b; Pavese *et al.* 2010b; Fellmuth *et al.* 2012; Pavese *et al.* 2013; Steur *et al.* 2015; Steur *et al.* 2017].
The triple points of $e$-$H_2$ and Ne are required in SPRT subrange 2, (25–273.16) K. The vapour pressure points at ≈17 K and ≈20.3 K of $e$-$H_2$ are required in the range (13.4–273.16) K. The use of the triple point of water (TPW) is prescribed for the whole part of the ITS-90 that is based on resistance- thermometer ratios, extending below 273.16 K down to 13.8 K and above 273.16 K up to the silver point, so affecting all comparisons including these ranges, based on the resistance ratios $W = R(T_{90})/R(\text{TPW})$.

---

[*] Associated scientist 2010-2015. Formerly, Consiglio Nazionale delle Ricerche, Italy, then INRIM until 2008.

In addition, the triple points of $e$-$H_2$ and Ne also affect the range covered by the interpolating gas thermometer (ICVGT), being two of the three fixed points of the ICVGT defined by the ITS-90 in the range 3–25 K—the third being the boiling point of $^4$He.

At the time when the key comparisons (KC) CCT-K1 "Realisations of the ITS-90, 0.65 K to 24.5561 K, using rhodium-iron resistance thermometers" (1997-2001) [CCT-K1], CCT-K2 "Key Comparison of capsule-type standard platinum resistance thermometers from 13.8 K to 273.16 K" (1997-99) [CCT-K2] and CCT-K7 "Key comparison of water triple point cells" (2002-04) [CCT-K7] were organised and completed the above issue was not yet recognised as important, so not yet formally included in the protocols. Subsequent CCT-K1.1 (2006-14, results not yet available according to the BIPM KCDB) [CCT-K1.1] and EUROMET.T-K1 (2008-12, similarly) [EUROMET.T-K1] did not take the isotopic effect into account. CCT-K2.1 (2003) [CCT-K2.1] and CCT-K2.4 (2006) [CCT-K2.4] did not take the isotopic effect into account; CCT-K2.3 (2006) [CCT-K2.3] *did* take the isotopic effects into account (official correction for $e$-$H_2$ and $H_2O$; VSL un-official evaluation for Ne, see later Footnote a of the Online Supplementary Information); also CCT-K2.5 (2015) [CCT-K2.5] *did* take the isotopic effects into account; CCT-K2.2 (2014) [CCT-K2.2], not yet completed, will also take the isotopic effects into account. The EUROMET.T-K7.1 (2008-09) [EUROMET-K7.1] and APMP.T-K7 [APMP.T-K7] included (optionally in the former) the isotopic issue in the comparison for water.

For water, the issue also involves the present definition of the kelvin, as modified in 2005 to include a reference isotopic composition [CI-2005]. In the ITS-90, for $e$-$H_2$ and $H_2O$, corrections to a reference composition were made formally available since the first version of the Technical Annex to the *Mise en pratique* of the kelvin in 2006; for neon it was since its 2014 version [MeP 2014].

At present, several cases are known of ITS-90 national realisations having adopted, at least partially, isotopic reference compositions: for example, NIST for the all ranges between 4 K and 273.16 K only for $e$-$H_2$ and of $H_2O$ [Tew, Meyer 2009]; INRIM only for the ICVGT for both $e$-$H_2$ and Ne [Steur, Giraudi 2013].

The study in this paper intends to provide evidence of the consequences of taking the isotopic effect into account. This is best done by using the outcomes of inter-comparisons, because one can also understand to which extent such a correction have affected, and will possibly affect, the differences between laboratories, when they were obtained in studies not having taken that effect into account. In particular, the scrutiny of key comparisons already available from the BIPM KCDB is important, because that MRA exercise provides to metrology the most valuable results, also in respect to the CMC declarations. However, this paper does not intend to tackle any formal consequence that may arise from, or be related to, the isotopic corrections.

In general, a study on the effect that the correction for the isotopic composition may have on the realisation of the ITS-90 in each laboratory is worthwhile if three conditions are met:
1. The isotopic composition of the samples used in a comparison are known;
2. The equation to compute the temperature correction is included in the current Annex to the kelvin MeP;
3. The correction can be applied to the results of a substantial number of participants to the comparison.

*Hydrogen*. The present information concerning the correction of hydrogen and for CCT-K1 and K1.1 comparison is quite limited, so the third condition is not met. In addition, the effect of the correction on the latter is almost irrelevant with respect to the comparison uncertainty. Similarly, for the CCT-K2.(x) the third condition is not met.

*Water*. The effect of water isotopic composition will not be analysed in this paper, being minimal in the temperature range below 25 K.

*Neon*. For neon it is possible to assign the isotopic composition to the gas samples used in a few open-cell realisations or contained, in most cases, in permanently-sealed cryogenic metal cells [Pavese *et al*. 2013]. In these cases, it is possible to apply the equations in the ITS-90 Technical Annex [MeP 2014] and compute the results at the reference composition. For neon all the above conditions are met for the CCT-K2, K2.1, K2.3 to K2.5.
In addition, some data are also traceable to the first International Intercomparison of sealed cells performed in 1978-84 [Pavese *et al*. 1984] or also ensures traceability for several results of the 1997-2005 Star Intercomparison [Fellmuth *et al*. 2012].

Therefore, this paper limits to neon the computation of the corrections and the discussion of some consequences, as an example of the complexity of the information needed to perform sound corrections, which may also affect the same type of corrections for other substances. See [Pavese 2014] for the way the information drawn from [MeP 2014] should be used to take isotopic composition into account in the calibration of SPRTs on the ITS-90, and [Steur *et al*. 2017] for details about the needed isotopic-composition assays and their outcomes.
In an Appendix, the effects of the chemical impurities in neon are also briefly presented, presently *not* subject to correction but only considered as an uncertainty component, to compare the importance of their effect with the isotopic effect.

**2 Isotopic effects on ITS-90 for the neon triple point temperature (24.5561 K)**

During a worldwide study lasted about 10 years, 26 different bottles of neon of commercial origin, plus three certified reference mixtures, were studied, including isotopic composition and chemical impurities assays, and thermal studies were performed on 34 samples drawn from them [Pavese *et al*. 2013]. These studies and the subsequent ones on pure $^{20}$Ne and $^{22}$Ne samples [ Pavese *et al*. 2013] led to the equation, now included in [MeP 2014], relating $T_{90,\mathrm{ref}}$ (ITS-90 defined value) to the value $T_{90}$ for the isotopic composition of the sample used, and allow to compute, from the measured resistance-ratio value, the corresponding value at $T_{90,\mathrm{ref}}$ [Pavese 2014].

In Table 1 the data are reported for the outcomes of several comparisons concerning neon, and in Table 2 the results for the CCT-K2.x of having taken into account the isotopic effect, based on the assay values selected after the critical evaluation of the assays, and their associated uncertainties [Pavese *et al*. 2013, Steur *et al*. 2017].[1] In Table 4 the results of the isotopic corrections for the Star intercomparison are reported [Fellmuth et al. 2012]. For important specific conditions concerning the way the data of each laboratory were obtained, see the Online Supplementary Information (IOT) associated with this paper.

2.1 Taking the effect of the isotopic composition into account
We recall here that, according to the MRA, the KCRV of the comparison CCT-K2 is common to all the subsequent integrations of its results with the results of the subsequent supplementary comparisons. It is not affected by uncertainty in [CCT-K2].
In order to take into account the effect of the isotopic composition on $T_{\mathrm{tp,Ne}}$, it is useful to summarise the exact meaning of the CCT-K2 results (*not including the CCT-K2.x*), and the procedure for applying the isotopic correction to them:

---

[1] All uncertainties *u* in this paper are the standard deviations ($k = 1$); *U* is the expanded uncertainty ($k \approx 2$).

a) Each participant used a sample of neon whose effect of the isotopic composition, at that time, was taken into account in the uncertainty budget only. This contribution to uncertainty is reported in Table 3, whose mean value amounts to 305 µK out of 361 µK of the total mean laboratory budget (85%) and out of 517 µK of the total comparison mean budget (59%)—so being the dominant contribution;

b) The results of the realisation of the triple point temperatures were compared through exchange of thermometers calibrated without taking into account the isotopic effect. However, being the triple point of neon a fixed point of the ITS-90, each participant laboratory associated to the provided measured value of the resistance ratio $R_{tp,Ne}/R_{TPW}$ the ITS-90 temperature value, 24.5661 K, exact. When the thermometers were compared in a comparison block at NRC, the measured resistance ratios did not exactly reproduce the supplied values—being that evidence the very reason of the comparison;

c) According to the CCT-K2 protocol, though one cell, NRC F15, was taken as the reference, the value 24.5561 K was not associated to it as the KCRV of the comparison. Instead, the resulting differences in the results, expressed as $\Delta T_{meas}$, were computed in [CCT-K2] with respect to a $T_{KCRV}$ being the *weighted mean* of the resulting temperatures;[2]

d) Normally those differences would directly express the difference in the realisations of the fixed point between the participant laboratories, $T_{thermal}$, due to thermal or technical effects. However, in this case, the measured temperatures were instead $T_{meas} = T_{thermal} + DT$, where:

(c1) a $DT_x$ is the temperature difference due to the isotopic composition of a sample with respect to the reference composition defined after 2014, the $^{IUPAC}x(Ne)$ one. Thus the corrections $DT_x = T_{meas,x} - 24.5661$ K;

(c2) all the remaining items of the uncertainty budget that are usual in a comparison, are taken into account for $T_{thermal}$. Notice that the $KCRV_{bc}$ used in [CCT-K2] is affected by the $DT_x$—see item f).[3] Thus, $T_{thermal} = T_{meas} - DT_x = T_{meas} - (T_x - 24.5661 \text{ K}) = 24.5661 \text{ K} + (T_{meas} - T_x)$. However, the final aim of this paper is instead to find $\delta T_{thermal} = T_{thermal} - KCRV_{ac}$.

e) Let us start from the fact that $\Delta T = T_{meas} - T(KCRV_{bc}) = T_{meas} - \text{wmean}(T_{meas})$. This can be approximated by replacing the weighted mean with the simple mean: $\Delta T = T_{meas} - \text{mean}(T_{meas}) = T_{meas} - \text{mean}(T_{thermal}) - \text{mean}(DT) = T_{thermal} + DT - \text{mean}(T_{thermal}) - \text{mean}(DT)$.

f) Then, one can compute the net contribution for each sample:
$T_{thermal,x} = \text{mean}_{ac}(T_{thermal}) + \Delta T_{meas,x} - DT_x + \text{mean}(DT_x)$, where the last term takes into account the offset in the original $KCRV_{ac}$, (1)
and finally, $\delta T_{thermal} = [\text{mean}_{ac}(T_{thermal}) + \Delta T_{meas,x} - (DT_x - \text{mean}(DT_x))] - \text{mean}(T_{thermal})$;
$\delta T_{thermal} = \Delta T_{meas,x} + (\text{mean}(DT_x) - DT_x)$. (2)

The method used in this paper aims at implementing the above procedure based on temperature values. First, one needs to compute the value of $KCRV_{bc}$, not explicitly reported in [CCT-K2].

*2.1.1 Main comparison (CCT-K2)*

The comparison did not define a "reference cell" to which assign the ITS-90 value, 24.5561 K but, as recalled above, the temperature value of the KCRV of CCT-K2 was computed as the *weighted mean* of the temperature values measured in the comparison block by each calibrated thermometer participating in the comparison, leading to the $\Delta T$s values in Table 1: the value of $T_{90,K2}$ assigned to the KCRV was not indicated in the Final Report.

---

[2] In this paper, the Greek $\Delta$ is used for differences *before* isotopic correction (e.g., $\Delta T_{meas} = \Delta_{or}$ in Table 1), while capital Roman D is used for the isotopic effect—see text in d). In this paper the differences due to a different amount of chemical impurities is not considered—see the Appendix.

[3] In this paper subscripts bc—for before correction—and ac—for after correction—are used. Thus the KCRVs are indicated in the following as $KCRV_{bc}$ and $KCRV_{ac}$, respectively.

arXiv:1704.05054v2 170421v2

When instead the isotopic composition is taken into account, *an arbitrary choice for $T_{KCRV}$ is not allowed anymore*, since the ITS-90 definition was later integrated by attributing the value 24.5561 K to, and only to, neon having the reference isotopic composition, the one recommended by IUPAC, $^{IUPAC}x$(Ne): $^{22}x = 0.0925$; $^{21}x = 0.0027$; $^{20}x =$ the rest [IUPAC].

This means that, in principle, the CCT-K2 KCRV *after* correction *is unlikely to be equal to* the CCT-K2 KCRV *before* correction, i.e. to the one used to express the differences in Table 1.

The $T_{90}$(KCRV$_{ac}$) and difference (KCRV$_{ac}$ – KCRV$_{bc}$) can now be evaluated with good approximation. Should the KCRV be the simple mean of the $T_{meas}$, it would be exact to say that KCRV$_{ac}$ = KCRV$_{bc}$ + mean(D$T_x$); in this case it is a good approximation because the corrections are small with respect to the temperature values. In addition, as illustrated in Section 2.1, one is not interested in the KCRV$_{ac}$($T_{meas}$), as it would directly come from the elaboration of the Final Report of CCT-K2, but in the KCRV$_{ac}$($T_{thermal}$), i.e. based on the measured values cleaned from the isotopic effect, $T_{thermal} = T_{meas} - DT_x$.

Being not all corrections necessarily exactly consistent with each other, the resulting value of the KCRV$_{ac}$ can vary somewhat depending on the correction chosen as the reference (exact) one.

In order to first obtain the value of the KCRV$_{bc}$($T_{meas}$), the method used in this paper is the following (where #1 and #2 indicate the thermometer set): [4]

i) The value $T_{90}$(Ne) = 24.5561 K, exact, corresponds to $^{IUPAC}x$(Ne);
ii) A *reference sample is chosen*. The choice of the **NRC F15** sample seems the most obvious, since NRC was the pilot in all K2.x comparisons;
iii) For NRC's *last* reference cell, Cu-M-1, the isotopic-effect difference to $^{IUPAC}x$(Ne) is D$T_{Cu-M-1}$ = –6(94) μK;
iv) Thus, the ITS-90 value of the NRC Cu-M-1 cell is **$T_{90}$(Cu-M-1)$_{ac}$ = 24.556 09$_4$ K**;
v) The NRC difference *measured* through cell F17 [$T$(Cu-M-1) – $T$(F15)]$_{bc}$ = –165(200) μK, so one gets **$T_{90}$(F15)$_{ac}$ = 24.556 25$_9$ K**;
vi) The differences $\Delta T_{F15}$ indicated in [CCT-K2] are $\Delta T_{F15\#1} = T$(F15 – KCRV)$_{\#1}$ = –0.06(44) mK and $\Delta T_{F15\#2} = T$(F15 – KCRV)$_{\#2}$ = –0.12(44) mK;
vii) Thus **$T_{90}$(KCRV$_{bc}$)$_{\#1}$ ≈ 24.556 32 K** and **$T_{90}$(KCRV$_{bc}$)$_{\#2}$ ≈ 24.556 38 K;**
viii) Incidentally, the isotopic-effect difference *from the assays* is [$T$(Cu-M-1) – $T$(F15)] = –342(95) μK: this is not a discrepancy since it is a different component of the cell differences.

Figure 1 depicts graphically the above procedure.

The *temperatures* actually measured during the CCT-K2, $T_{meas}$, are obtained by adding to $T_{KCRVbc}$ the $\Delta T_{meas}$ values recorded under "Results" in [CCT-K2] for each sample.

One could then compute the $T_{meas,ac}$ by simply adding to $\Delta T_{meas}$ the D$T_x$ obtained from the ITS-90 Technical Annex of [MeP], and then compute the weighted mean from the latter set, for both sets #1 and #2: $\delta T_{meas,ac} = T_{meas,corr} - T$(KCRV$_{ac}$). The isotopic corrections are reported in Table 2 in the column "Isotopic D$T$", For the isotopic composition of the samples, see [Pavese *et al.* 2013; Steur *et al.* 2015; Steur *et al.* 2017]. The KCRV$_{ac}$ are reported in Table 2: $T_{KCRVac}$ = 24.556 47$_1$ K for thermometers #1, and $T_{KCRVac}$ = 24.556 55$_8$ K for thermometers #2, different, as expected, from the KCRVs before correction: notice that these values correspond to the values in item (viii) above well within the uncertainties. That change alone entails changes of +0.15 mK and +0.18 mK, respectively, to all the $\Delta T_{meas} = T_{90bc} − T_{KCRVbc}$ in Table 2—and in the Sections 2.1.4 to 2.1.6—but note that pair differences are unaffected (pair DoEs, see the IOT).[5]

---

[4] These values, as all the $\Delta T_{meas}$, are affected by the lack of isotopic correction.

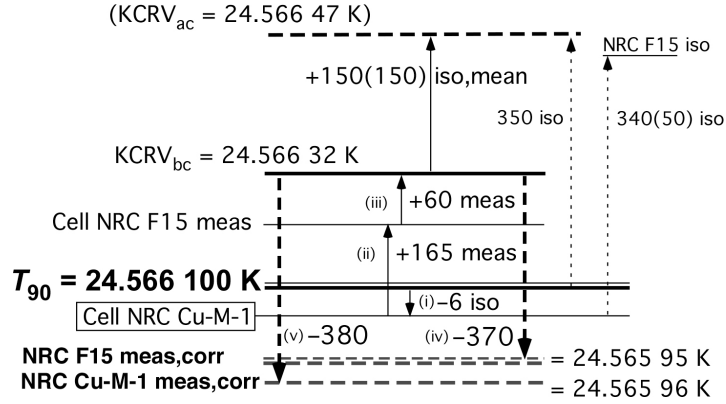

**Fig. 1.** Graphical representation of the procedure described in Section 2.1.1 for set #1. The procedure starts from cell NRC Cu-M-1, step (i), where $T_{ref} = T_{90,Ne} = 24.5561$ K. For the $KCRV_{ac}$ see Table 2 and Fig. 2. The $KCRV_{ac}$ is 24.566 47 K.

However, the above computation shows a limited interest, since the $T_{meas}$ are those biased by the isotopic effect through $\Delta T$. They should be transformed into the $T_{thermal}$, according to the procedure indicated in Section 2.1, an approximated one by using the simple mean of the $T_{meas}$.
Starting from Eq. (1) in Section 2.1 (f), the values known in it are those for: all $\Delta T_x$ from [CCT-K2] and all $DT_x$ from [MeP, Technical Annex]. Note that Eq. (1), *does not contain any absolute value* of $T$, but only mean or relative values: however, one obtains the temperature values in Table 2 as $T_{thermal} = 24.566100$ K $+ \delta T_{90,thermal}$. The $\delta T_{90,thermal}$ after correction replace the $\Delta T_{meas}$ before correction.
 The summary of the uncertainties is reported separately in Table 3—and commented in Section 2.2.

It is interesting to compare the $\delta T_{90,thermal}$ with the $\delta T_{meas,ac}$ computed before. Both are approximated: the latter because, as said, they use $T_{meas}$; the former because the simple mean replaced the weighted mean and they still use the $\Delta T$. However, the difference between the two is fixed and only +40 μK for #1 and +95 μK for #2. The reason is that, in fact, $\delta T_{90,thermal} − \delta T_{meas,ac} =$ $KCRV_{bc} − 24.5661$ K $−$ mean$(DT_x)$.
To be noticed that, after correction for the isotopic effect, the NRC experimental difference (Cu-M-1 −F15)$_{NRC}$ = −165(200) μK becomes (Cu-M-1 −F15)$_{thermal}$ +147(220) μK. However, this change does not require a correction in the procedure Section 2.1 (v) nor an iteration of the calculations, since in (v) one must use the KCRV based on which the values of the $\Delta T_{or}$ in Table 1 were computed, as taken from [CCT-K2].

*2.1.2 Comparison K2.1 (VNIIFTRI, NRC)*
In this comparison, the NRC reference cell was still F15. The isotopic composition of the VNIIFTRI sample used in the CCT-K2 is unknown, so no computation is possible to take it into account. Therefore, the measured differences +0.28 mK (#1) and +0.22 mK (#2) remain unchanged.

---

[5] The new KCRVs were obtained by omitting the INM datum, probably already omitted from the KCRV computation by NRC in the Final Report, and by including KRISS, whose datum was not processed in the Final Report [CCT-K2].

Should one assume that the sample in question is from the same bottle that was used for the cell participating to the 1978-84 Inter-comparison [Pavese *et al.* 1984] and the more recent Star intercomparison [Fellmuth *et al.* 2012], an isotopic correction of –0.29 mK would apply, leading to a difference of –0.01 mK (#1) and –0.07 mK (#2), respectively.

*2.1.3 Comparison K2.3 (NMI-VSL, NRC)*

In this comparison, the NRC reference cell was changed to the newest Cu-M-1, whose *uncorrected* difference from cell F15 has been measured at NRC (though cell F17) to be $[T(\text{Cu-M-1}) - T(\text{F15})]_{bc}$ = –165(200) μK. See Table 2 for the values before and after correction of $T_{90}(12\text{Ne})$, $T_{90}(\text{F15})$ and $T_{90}(\text{Cu-M-1})$.

NMI-VSL used INRIM cell 12Ne (5N gas sample from Messer Griesheim, with assay #11, [Steur *et al.* 2017] assigned isotopic correction 123 μK). Thus, from Table 2 the values after correction are $[T(12\text{Ne}) - T(\text{Cu-M-1})]$ = –0.000 55 K, $[T(12\text{Ne}) - T(\text{F15})]$ = –0.000 40 K and $[T(12\text{Ne}) - T_{90}]$ = –0.000 55 K. [6]

*2.1.4 Comparison K2.4 (INTiBS, LNE-INM, NRC)*

In this comparison, the NRC reference cell was also the newest Cu-M-1—see comparison K2.3.

INTiBS used INRIM cell E3Ne (5N gas sample from Messer Griesheim, with assay #11, assigned isotopic correction 123 μK). Thus, from Table 2 one gets the value of $T_{90}(\text{E3Ne})$ and the values after correction are $[T(\text{E3Ne}) - T(\text{Cu-M-1})]$ = –0.000 15 K, $[T(\text{E3Ne}) - T(\text{F15})]$ = 0 K and $[T(\text{E3Ne}) - T_{\text{KCRV}}]$ = –0.000 15 K. [7]

LNE-INM used cell Ne02/1 (5N gas sample from Air Liquide, with assay #14, assigned isotopic correction –32 μK). Thus, from Table 2 one gets the value of $T_{90}(\text{Ne02/1})$ and the values after correction are $[T(\text{Ne02/1}) - T(\text{Cu-M-1})]$ = –0.000 42 K, $[T(\text{Ne02/1}) - T(\text{F15})]$ = -0.000 27 K and $[T(\text{Ne02/1}) - T_{\text{KCRV}}]$ = –0.000 42 K. [7]

*2.1.5 Comparison K2.5 (NMIJ-AIST, INRIM, NRC)*

This comparison is the *only one* supplied with the results corrected for the isotopic composition of the samples. This requires an inverse computation in order to get the values before correction. For this comparison, the NRC reference cell was also the newest Cu-M-1—as with comparisons K2.3 and K2.4.

NMIJ-AIST used its cell Ne-5 (5N gas sample from AirWater, with assay #7, assigned isotopic correction 4 μK). Thus, from Table 2 one gets the value of $T_{90}(\text{Ne-5})$ and the values after correction are $[T(\text{Ne-5}) - T(\text{Cu-M-1})]$ = –0.000 32 K, $[T(\text{Ne-5}) - T(\text{F15})]$ = –0.000 18 K and $[T(\text{Ne-5}) - T_{\text{KCRV}}]$ = 0.000 32 K. [7]

INRIM used cell Ec2Ne (5N gas sample from Messer Griesheim, with assay #11, assigned isotopic correction 123 μK). Thus, from Table 2 one gets the value of $T_{90}(\text{Ec2Ne})$ and the values after correction are $[T(\text{Ec2Ne}) - T(\text{Cu-M-1})]$ = 0.000 59 K, $[T(\text{Ec2Ne}) - T(\text{F15})]$ = 0.000 44 K and $[T(\text{Ec2Ne}) - T_{\text{KCRV}}]$ = 0.000 59 K. [7]

2.2 Uncertainty of the CCT-K2 comparisons

The uncertainty issue has been treated separately in Table 3, since its complex analysis requires a full table.

Table 3 shows an important issue: every comparison exercise adds uncertainty to the previous results, in average a 30% more when comparing $U_{\text{KC}}$ to $U_{\text{TOTlab}}$. In addition, as expected, the increase is larger for the late K2.1 to K2.5 (≈30%) than for the original K2 (≈20%).

---

[6] The above values derive from considering the NRC F15 as the reference cell for the *original* CCT K2. The $\text{KCRV}_{\text{K2.x}}$ remains that of the CCT-K2.

Another very important issue is that, by strongly decreasing the uncertainty on the isotopic composition, one strongly affects the overall *laboratory* uncertainty budget of the comparison of neon samples: in fact the average contribution of the isotopic effect is of 305(97) μK out of a total of 361(145) μK, so accounting for the 85%.

Since the isotopic uncertainty drops in average from 305(97) μK to 37(33) μK, the laboratory differences decrease by about 30% in average after compensating for the isotopic effect, and the comparison uncertainty accordingly: *the benefit of the corrections amounts in average to* 60(15)%, i.e. it cuts the comparison uncertainty by more than half.

## 3 Discussion and Final Remarks

Figure 2 shows a graphical representation of data in Table 2: the mean value of the original deviations $\Delta T_{or}$ is –147(268) μK[7] for set #1 and –166(309) μK for set #2, while those after correction, $\delta T_{iso} = (T_{90ac} - KCRV_{ac})$, are –175(306) μK for set #1 and –187(388) μK for set #2.

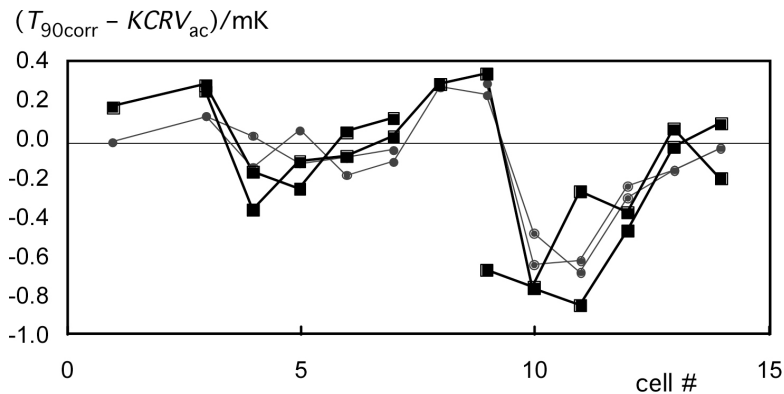

**Fig. 2.** Graphical representation of the corrected data $(T_{90ac} - KCRV_{ac}) = (T_{meas,ac} - T_{KCRVac})$ from Table 2 and the uncorrected data $\Delta T_{original}$ from Table 1, for cells #1 to #15 and for thermometer sets #1 and #2. Gray dots and lines: uncorrected differences. Black squares and lines: isotopic-corrected differences.

However, when subtracting from the original differences the contribution of the isotopic effect, in Fig. 4 one gets for $\delta T_{thermal,ac}$ –167(233) μK for set #1 and –147(240) μK for set #2, where the uncertainty is reduced by 60% in average, as already observed in Table 3. Actually, apart two cells, after correction the differences are within the interval (+0.3, –0.2) mK, while in Fig. 2 they were in the interval (+0.4, –0.8) mK.

Therefore, by taking into account the isotopic effect, one can have a substantial improvement in the quality of the comparison results of the CCT-K2.x, though the uncertainty will increase progressively with time for the supplementary comparisons on the same fixed point—see Table 3 and Section 2.2.

In some cases it is possible to compare cells differences of INRIM production or of cells of other NMIs *directly* measured also at INRIM [Pavese *et al.* 2010b] with the values obtained from the K2.x ones.

In Fig. 3 the following cells are shown, all sealed with gas taken from the same bottle of gas (Messer Greisheim, analysis #11 [Steur *et al.* 2017]): from the left, INTiBS (INRIM); cell E3Ne [CCT-K2.4], that was made in the *same* batch (24 Aug 2000) of cell E4Ne; VSL (INRIM) cell

---

[7] In parenthesis the standard deviation.

12Ne [CCT-K2.3] that was made in the *same* batch (21 Oct 1999) of cell 15Ne; PTB (INRIM) E1Ne (12 Dec 1999) [Fellmuth *et al.* 2012] that was sealed two months before the E2Ne to E4Ne batch; (lower data) INRIM Ec2Ne reference cell as measured in 2008 [Pavese *et al.* 2010b]; finally, (upper data) INRIM Ec29Ne that was sealed from the same bottle of gas on its return back to INRIM after the assays at IRMM, and measured in 2015. All results are compatible with each other except the last one with respect to the 2008 ones. The +94 µK increase of $T_{tp}$ is attributed in [Steur *et al.* 2017] to a possible change for unknown reasons of gas isotopic composition in the bottle during the years.

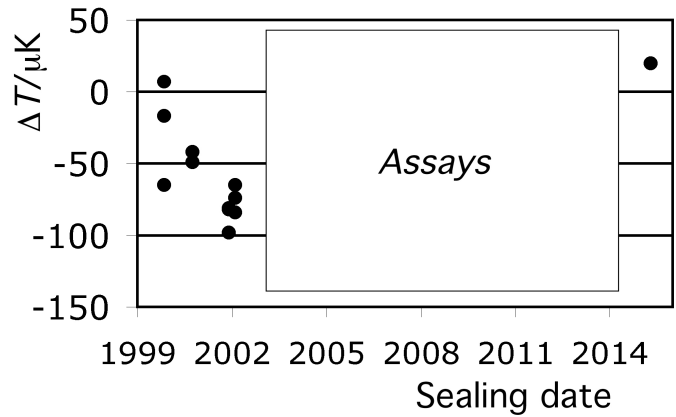

**Fig. 3**. Differences between samples drawn at INRIM at different times (from left to right) from the *same bottle*; zero of $\Delta T$ arbitrary, hydrostatic head correction applied. Sealing dates of INRIM cells: from left, cell 12Ne–15Ne; cell E2–E4Ne; cell Ec1Ne; cell Ec2Ne sealed in 2003 as measured in 2008; cell Ec29Ne sealed and measured in 2015. Uncertainty of each determination is ± ≈50 µK.

In a *direct* cell comparison under European Project MULTICELLS [Pavese *et al.* 2003], differences were found as follows, with respect to cell 7Ne, supplied by INRIM —sealed in 1986 and the batch 6Ne - 11Ne, the same of cell #20 in this paper): INRIM Eb1Ne 0.024(18) mK; [8] INRIM Ec2Ne – 0.065(17); LNE-INM Ne99/2 = 0.31(20) mK. Under the same Project, at VSL the difference (LNE-INM Ne02-1 – INRIM 3Ne) was found to be –0.07(45) mK.

To complete the results from previous inter-comparisons, the bilateral DoE of the Int84, as reported in [Pavese *et al.* 1984], are the following with respect to IMGC-CNR reference cell—where the difference between cells 1Ne and 3Ne was then set to 0.00 mK: ASMW (later PTB) 0.34 mK, INM –0.03 mK, NRC –0.04 mK, NRLM = –0.13 mK and PRMI (VNIIFTRI) 0.06 mK; $U$ was estimated to be 0.3 mK. INM and VNIIFTRI cells are traceable to present data.

The results of the CCT-K2.x can be compared with the results of the largest *direct* comparison of samples in sealed cells made after the Int84 [Pavese *et al.* 1984]: the Star comparison [Fellmuth *et al.* 2012], whose data are compared in Table 4 using the data of Table 1.
Figure 4 (right part) makes the improvements self-evident with respect to the K2.x data (left part). Only two samples are outlying: INM Ne02/1 and NIST 201. The latter is greyed in Table 4 because traceability back to the right filling gas is unsure. With its exclusion, the mean of the corrected differences is 56(68) µK (74(87) µK before isotopic correction), with a measurement uncertainty ($k$ = 1) of ≈47 µK, thus not significant at the $U$ level. Except one, all deviations are within ± 50 µK.

---

[8] Here the uncertainty in parenthesis is the expanded one, $U$, which is the overall measurement uncertainty budget, so including the isotopic effect.

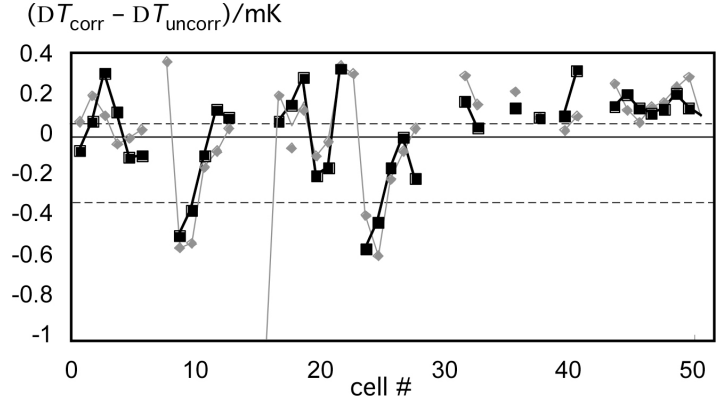

**Fig. 4**. Graphical representation of differences from the KCRV of K2-xx and STAR direct-cell comparisons, uncorrected (gray dots and lines) nd corrected (black squares and lines):
$(T - KCRV_{K2ac}) = DT_{thermal,ac}$ (Table 2); $(T - KCRV_{K2bc}) = \Delta T_{original}$ (Fig. 1).
• **On the left until #28**: **K2-xx** differences for cells #1 to #15 and thermometer sets #1 and #2. $\Delta T_{original,\#1} = -147(268)$ μK, $\Delta T_{original,\#2} = -166(309)$ μK; $DT_{thermal,ac,\#1} = -167(233)$ μK, $DT_{thermal,ac,\#2} = -147(240)$ μK.
 **On the right from #30 to end**: **STAR** differences (Fig. 4) (56(68) μK —74(87) μK before isotopic correction), $u = 47$ μK. [Fellmuth *et al.* 2012]
The dotted lines indicate the range of the isotopic effect for the studied samples, as obtained from the MeP in Table 2.

A similar comfortable situation was found in the more restricted *direct* inter-comparison of sealed cells performed at INRIM [Pavese *et al.* 2010b]:
(PTB Ne-7 – INRIM Ec2Ne) = 0.057(52) mK; [9]
(NMIJ Ne-2 –INRIM Ec2Ne) = 0.006(57)—whence (NMIJ Ne-5 –INRIM Ec2Ne) = 0.043(57);
(NPL Ne2 – INRIM Ec2Ne) = 0.154(53) mK;
(INTiBS E3Ne – INRIM Ec2Ne) = 0.029(54) mK (from Fig. 2);
(VSL 12Ne – INRIM Ec2Ne) = 0.049(54) mK (from Fig. 2).
Further, (15Ne – Ec2Ne) = 0.054(58) mK; (1Ne –Ec2Ne) = 0.131(66).

Furthermore, using cell INRIM Ec2Ne as reference, via PTB Ne-12 by knowing that it is 0.008 mK hotter than PTB cell Ne-7 as measured in [Fellmuth *et al.* 2012] where the latter is the reference, the differences to PTB Ne-7 are:
(3Ne – Ne-7) = –0.066(45) mK and  (12Ne –Ne-7) = 0.062(48) mK, respectively. [9] In addition, (E1Ne –Ne7) = –0.023(45) mK.
Thus, being PTB Ne-12 hotter than INRIM Ec2Ne by 0.054(33) mK—so (Ne-7 – Ec2Ne) = 0.046 mK—one finds (E3Ne – Ec2Ne) = –0.031 mK and (12Ne –Ec2Ne) = 0.108 mK, respectively, and (E1Ne – Ec2Ne) = 0.069 mK.

**APPENDIX A. Chemical impurities**

The *chemical impurities*, reported in the column of chemical corrections of Table 2, show that: [9]

– the corrections for these impurities may be even more critical that the ones for the isotopic composition. The experience of the International Project on isotopic neon [Pavese *et al.*

---

[9] Effect on $T_{tp,Ne}$: for $H_2$ (–7 ± 3) μK/$10^{-6}$; for $N_2$ (–8 ± 3) μK/$10^{-6}$.

2013] has shown that the availability from KRISS of excellent assays also on the chemical impurities was a basic asset to obtain an outstanding top-accuracy *overall* correction of the results on the measured samples in that Project;
- The lack of data on some critical impurities, namely like hydrogen in neon, can impair the validity of the correction for chemical impurities—and the usefulness of the isotopic correction. [Steur *et al.* 2017]

It is the authors' opinion that the CCT should consider the need, for top accuracy, of the application of corrections, by using the SIE or similar methods [Fellmuth *et al.* 2015; Pavese 2009; Pavese 2011].

**APPENDIX B. Specific conditions for each laboratories concerning the K2.(x) comparisons**

**BNM-INM** (now **LNE-LCM**)—The sample BCMH2O of neon used in K2 is *not* traceable to isotopic assays. However, from the STAR comparison [Fellmuth *et al.* 2012] it can be inferred that the bottle was the same as for INM sample 99/2, for which an isotopic assay is available [Steur et al. 2017].
Later another measurement was available in the frame of the comparison K2.4 in 2005-2006, using cell Ne02/1 (see Table 1 in the paper also for the next references to samples), for which an isotopic assay exist: this result is also reported.
Finally, the BNM-INM datum in K2 for set #2 was a clear outlier, though within its very large uncertainty.

**CNR-IMGC** (now **INRIM**)—The sample of neon used in K2 was sealed in cell 3Ne, later outsourced by the IMGC to NIM, and the IMGC reference was later moved to cell 1Ne, using a different bottle of gas (see Table 2 in the paper). However, as explained in [Pavese *et al.* 2010b] and its online Supplementary Information, the thermal results obtained with cells 1Ne and 3Ne are not consistent with the different measured compositions: the difference in the isotopic correction according to the assays amounts to (1Ne − 3Ne) = 195(30) μK. The difference ($T$1Ne − $T$3Ne) was –44(110) μK in INRIM measurements [Pavese *et al.* 2010b] and −29(76) μK in [Fellmuth *et al.* 2012]—the latter will be used in the following. On the other hand, the isotopic assays have some reasons for being less accurate than claimed because the sample available for the assays was quite small: the bottles only contained 2 mmol and 3 mmol, respectively, of gas, decades old. In addition, in the set of observed $T$L $vs$ 22$x$ values, the $T$L(3Ne) looks as an outlier being too low by 100-150 μK, consistent with the above discrepancy. In Table 2 of the paper the values for sample 1Ne are also provided.
INRIM performed a comparison of some key cells in 2008-09 [Pavese *et al.* 2010b]. This is referred in Table 2 as "INRIM" comparison, having INRIM cell Ec2Ne as the reference.

**KRISS**—This laboratory was not considered with respect to the CCT-K2 comparison until recently [Yang *et al.* 2015], when the isotopic composition of the sample of neon used for the (open cell) realisation during CCT-K2 became available. In addition, in [Yang *et al.* 2015] the realisation of the triple point of neon is also reported using samples from three different gas bottles whose isotopic composition is known from a *calibrated* mass spectrometer.

**NBS** (now **NIST**)—The sample of neon used for the (open cell) realisation during CCT-K2 was taken from a known bottle, whose isotopic assay was made available in 2003. For NIST, it is known that the same gas was used also for the data of CCT-K1 by direct realisations of the ITS-90 using an interpolating constant-volume gas thermometer.

**NPL**—The sample of neon used for the sealed cell realization during CCT-K2 for thermometer 1728839 was sealed in the NPL cell Ne2 and Ne1: NPL assumed the two cells to having been filled from the same bottle, for which isotopic assays became available in 2003. The data for thermometer 213865 came instead from calibration from comparison with the NPL reference thermometer.

**NRC**—The sample of neon used for the sealed cell realization during CCT-K2 was sealed in the NRC cell F15, taken from a bottle whose isotopic assays became available only in 2003. The same bottle is said by NRC to have been used for sealing cell F17. See hereinafter and the main text. [CCT-K2.3; CCT-K2.4; CCT-K2.5]

**PTB**— The sample of neon used for the sealed cell realization during CCT-K2 was sealed in the PTB cell Ne-7, taken from a bottle for which isotopic assays became available in 2003. The PTB cell Ne-12, sealed later from the same bottle, is known to be basically identical to cell Ne-7, $^{TM}T = 8(66)$ μK. Only thermometer 1842379 was used for the K2 equivalence table.

PTB was the co-ordinator of a subsequent comparison [Fellmuth *et al.* 2012], reported in Table 2 as Star intercomparison. In this paper, INRIM cell Ec2Ne is used as the reference cell also for this exercise.

**PRMI** (later **VNIIFTRI**)—The sample of neon used for participation to the CCT-K2 is unknown. In all instances, the PRMI withdrew from the K2. Later, a bilateral comparison, K2.1, was performed with NRC. The pilot still used a thermometer calibrated on cell F15, so that no calculation was necessary to refer to the CCT-K2 KCRV. The VNIIFTRI thermometers were calibrated against an *average* realisation of the ITS-90, so that no data on the sample of neon used are available. No correction for isotopic composition can be performed.

In 1978-84, the PRMI participated in the first International Intercomparison of fixed points in sealed cells [Pavese *et al.* 1984] with sealed cell MC-897, whose relationship with other still-existing sealed cells exist, and quite recently it was re-measured in the frame of the Star intercomparison [Fellmuth *et al.* 2012], where it was found (MC-897 – 3Ne) = 90(92) μK. The isotopic composition of the gas used to fill this cell was provided by the VNIIFTRI with low resolution, corresponding to an uncertainty component of 80 μK.

**NMI** (then **NMI-VSL**, now **VSL**)— NMI withdrew from its initial participation to the CCT-K2. Later a bilateral comparison, K2.3, was performed with NRC. VSL used a sealed neon cell produced by IMGC, 12Ne. NRC was unable to still use a calibration on the cell F15 used at the time of CCT-K2, and had to use a chain of calibrations to relate the last cell used, Cu-M-1, and F15, through cell F17 (53 μK colder than F15). The resulting difference resulted to be (Cu-M-1 – F15) = –165(200) μK. Instead, the same difference obtained from the isotopic composition corrections is –342(95) μK (see Footnote a and more below under INTiBS for NRC). On the Report of comparison K2.3, VSL performed the isotopic corrections for *e*-H2 and H2O using the official equations [21] available in 2006. On the contrary, for Neon VSL used a home-made evaluation,
–157(7) μK; a for the latter now the official correction for the neon sample sealed in cell 12Ne is available and is: –123(20) μK. The difference with the VSL estimate is irrelevant considering the uncertainty of the results.

**INTiBS**—This Institute participated in a trilateral comparison, K2.4, with BNM and NRC (as the pilot). INTiBS used a sealed neon cell produced by the IMGC (E3Ne), with known isotopic composition. As in the case of K2.3, NRC was unable to still use a calibration based on the cell F15 used at the time of CCT-K2, and had to use a chain of calibrations to relate the last cell used, Cu-M-1, and F15, through cell F17. Assuming that cells F15 and F17 were effectively filled from the same neon bottle, the resulting difference resulted to be (Cu-M-1 – F15) = –165(200) μK.

**NMIJ-AIST**— This Institute participated in a trilateral comparison, K2.5, with INRIM and NRC, where the supplied temperature values are corrected for isotopic composition. Two thermometers were calibrated for this exercise, and the values are supplied for the neon isotopic composition corrected to the reference one [IUPAC], one sample being almost coincident with the latter. Only cell Ne-5 is the NMIJ-AIST reference, cell Ne-2 (see INRIM inter-comparison) being known to be hotter by 31(50) μK.

**References**

[APMP.T-K7]
http://kcdb.bipm.org/appendixB/KCDB_ApB_info.asp?cmp_idy=855&cmp_cod=APMP.T-K7&prov=exalead
[CCT-K1] CCT-K1 (NPL Pilot).
http://kcdb.bipm.org/appendixB/KCDB_ApB_info.asp?cmp_idy=453&cmp_cod=CCT-K1&prov=exalead
[CCT-K1.1] CCT-K1.1 (NIST (Pilot), NMIJ-AIST).
http://kcdb.bipm.org/appendixB/KCDB_ApB_info.asp?cmp_idy=794&cmp_cod=CCT-K1.1&prov=exalead
[CCT-K2] CCT-K2 (NRC Pilot).
http://kcdb.bipm.org/appendixB/KCDB_ApB_info.asp?cmp_idy=454&cmp_cod=CCT-K2&prov=exalead
[CCT-K2.1] CCT-K2.1 (NRC (Pilot), VNIIFTRI).
http://kcdb.bipm.org/appendixB/KCDB_ApB_info.asp?cmp_idy=697&cmp_cod=CCT-K2.1&prov=exalead
[CCT-K2.2] CCT-K2.2 (INRIM, NIM, NRC(Pilot)). Not completed yet
[CCT-K2.3] CCT-K2.3 (NMI-VSL, NRC (Pilot)).
http://kcdb.bipm.org/appendixB/KCDB_ApB_info.asp?cmp_idy=744&cmp_cod=CCT-K2.3&prov=exalead
[CCT-K2.4] CCT-K2.4 (INTiBS, LNE, NRC (Pilot)).
http://kcdb.bipm.org/appendixB/KCDB_ApB_info.asp?cmp_idy=745&cmp_cod=CCT-K2.4&prov=exalead
 [CCT-K2.5] CCT-K2.5 (INRIM, NMIJ-AIST, NRC (Pilot)).
http://kcdb.bipm.org/appendixB/KCDB_ApB_info.asp?cmp_idy=771&cmp_cod=CCT-K2.5&prov=exalead. K D Hill, T Nakano and P P M Steur, Metrologia 52 (2015) doi: 10.1088/0026-1394/52/1A/03003
 [CCT-K7] CCT-K7 (BIPM Pilot).
http://kcdb.bipm.org/appendixB/KCDB_ApB_info.asp?cmp_idy=459&cmp_cod=CCT-K7&prov=exalead
[ CI-2005] BIPM, Recommendation 2 (CI-2005): Clarification of the definition of the kelvin, unit of thermodynamic temperature. http://www.bipm.org/en/si/si_brochure/chapter2/2-1/kelvin.html
[EUROMET.T-K1] EUROMET.T-K1 (INRIM, INTiBS, NPL, PTB (Pilot), VSL).
http://kcdb.bipm.org/appendixB/KCDB_ApB_info.asp?cmp_idy=1290&cmp_cod=EURAMET.T-K1& prov=exalead (Draft A Report). C Gaiser, B Fellmuth, P P M Steur, A Szmyrka-Grzebyk, H Manuszkiewicz, L Lipinski, A Peruzzi, R Rusby, D Head, paper at TEMPMEKO 2016, to be published on IJOT [EUROMET-K7.1] EUROMET-K7.1 (NMI-VSL (Pilot), many others). http://kcdb.bipm.org/appendixB/KCDB_ApB_participant.asp?cmp_idy=789&cmp_cod=EUROMET.T-K7& prov=exalead

[Fellmuth *et al.* 2005] Fellmuth B., Wolber L., Hermier Y, Pavese F., Steur P.P.M., Peroni I., Szmyrka-Grzebyk A., Lipinski L., Tew W.L., Nakano T., Sakurai H., Tamura O., Head D., Hill K.D., Steele A.G., Metrologia 42 (2005) 171–193

[Fellmuth *et al.* 2012] Fellmuth B., Wolber L., Head D., Hermier Y., Hill K.D., Nakano T., Pavese F., Peruzzi A., Shkraba V., Steele A.G., Steur P.P.M., Szmyrka-Grzebyk A., Tew W.L., Wang L., White D.R., Metrologia 49 (2012) 257–265

[Fellmuth *et al.* 2015] Fellmuth B., Hill K.D., Pearce J.V., Peruzzi A., Steur P.P.M., Zhang J., Guide to the Realization of the ITS-90. Fixed Points: Influence of the Impurities. BIPM, CCT, update of 21 October 2015

[Hill 2013] Hill K.D., private communication (2013)

[ITS-90] Preston-Thomas H., The International Temperature Scale of 1990 (ITS-90) Metrologia 27 (1990) 3–10; Errata 27 (1990) 107

[IUPAC] Wieser M.E., Coplen T.B., Pure Appl. Chem. 83 (2011) 359–396

[MeP 2014] BIPM, *Mise en pratique* for the definition of the kelvin, 2014 http://www.bipm.org/en/si/si_brochure/chapter2/2-1/kelvin.html , and its Technical Annex

[Nicholas *et al.* 1996] Nicholas J.V., White D.R., Dransfield T.D., 1997 Proc. TEMPMEKO 1996 ed. P. Marcarino (Torino, Italy: Levrotto e Bella) pp 9–12

[Pavese *et al.* 1984] Pavese F., Ancsin J., Astrov D.N., Bonhoure J., Bonnier G., Furukawa G.T., Kemp R.C., Maas H., Rusby R.L., Sakurai H., Ling Shankang, Metrologia 20 (1984), 127–144. Full Report available as: Pavese F., Final Report, Monograph 84/4 of Bureau International des Poids et Mesures, BIPM Sèvres (1984), pp.219 (http://www.bipm.org/en/publications/monographies-misc.html). In the format of a CCT-KC: Pavese F., Comptes Rendus Comité Consultatif de Thermométrie, BIPM Sèvres (2000), Doc.CCT/2000-7

[Pavese, Tew 2000] Pavese F., Tew W.L., 2000 Comité Consultatif de Thermométrie (CCT) Working Document (Sèvres: Bureau International des Poids et Mesures) Doc. CCT/00–9

[Pavese *et al.* 2003] Pavese F., Fellmuth B., Head D., Hermier Y., Peruzzi A., Szmyrka Grzebyk A., Zanin L., Temperature, Its Measurement and Control in Science and Industry, D.Ripple ed., Amer.Inst.of Phys., New York A, 8 (2003), pp. 161–166

[Pavese 2005] Pavese F., Metrologia 42 (2005) 194–200

[Pavese *et al.* 2005] Pavese F., Fellmuth B., Head D., Hermier Y., Hill K.D., Valkiers S., Analytical Chemistry 77 (2005) 5076–5080

[Pavese 2009] Pavese F., Metrologia 46 (2009) 47–61

[Pavese *et al.* 2010] Pavese F., Steur P.P.M., Valkier S., Nakano T., Sakurai H., Tamura O., Peruzzi A., Fellmuth B., Wolber L., Lipinski L., Szmyrka-Grzebyk A., Tew W.L., Hill K.D., Steele A.D., Hermier Y., Sparasci F., Int. J. Thermophysics,31 (2010) 1633–1643

[Pavese *et al.* 2010b] Pavese F., Steur P.P.M., Bancone N., Ferri D., Giraudi D., Metrologia 47 (2010) 499–518

[Pavese 2011] Pavese F., Metrologia 48 (2011) 268–274

[Pavese *et al.* 2013] Pavese F., Steur P.P.M., Hermier Y., Hill K.D., Kim Jin Seog, Lipinski L., Nagao K., Nakano T., Peruzzi A., Sparasci F., Szmyrka-Grzebyk A., Tamura O., Tew W.L., Valkiers V., van Geel J., in Temperature, Its Measurement and Control in Science and Industry, Vol 8: Proceedings of the Ninth International Temperature Symposium, AIP Conf. Proc. 1552 (2013) 192–197; doi: 10.1063/1.4821378

[Pavese 2014] Pavese F., Int. J. Thermophysics, 35 2014 1077–1083

[Steur, Giraudi 2013] Steur P.P.M., Giraudi D., in Temperature, Its Measurement and Control in Science and Industry, Vol 8: Proceedings of the Ninth International Temperature Symposium, AIP Conf. Proc. 1552 (2013) 124–129; doi: 10.1063/1.4819526

arXiv:1704.05054v2 170421v2

[Steur *et al.* 2015] Steur P.P.M., PaveseF., Fellmuth B., Hermier Y., Hill K.D., Kim Jin Seog, Lipinski L., Nagao K., Nakano T., Peruzzi A., Sparasci F., Szmyrka-Grzebyk A., Tamura O., Tew W.L., Valkiers S., van Geel J., Metrologia 52 (2015) 104–110

[Steur *et al.* 2017] Steur P.P.M., Yang Inseok, Kim Jin Seok, Nakano T., Nagao K., Pavese F., An intercomparison of replicated isotopic-composition assays of neon and their relation to thermal analyses, TEMPMEKO 2016, Zakopane, September 2016. Submitted to Analytical Chemistry.

[Tew, Meyer 2009] Tew W.L., Meyer C.W., Doc. CCT/08-09, BIPM, Sèvres, 2009

[Yang *et al.* 2015] Yang I., Gam K.S., Joung W., Kim Y.G., Int. J. Thermophysics 36 (2015) 2072–2084

[White *et al.* 2003] White D.R., Dransfield T.D., Strouse G.F., TewW.L. , Rusby R.L., Gray J., 2003, in Temperature, Its Measurement and Control in Science and Industry vol 8, ed D Ripple (New York: AIP) pp 221–227

Table 1 – Data used in this study from comparisons involving the Neon triple point, $T_{90\text{tp}}$ = 25.5661 K. Basic data.

| # | Data from comparison [a] | Sample fabricated (owned by) | Year (S=sealed) | Cell # [b] | Gas (Assay. #) [c] | Purity, nom. (measured) | Set#1 CCT-K2.x comparison results [d] | | Set#2 | |
|---|---|---|---|---|---|---|---|---|---|---|
| | | | | | | | $\Delta T_{\text{or}}$/mK | $U$/mK | $\Delta T_{\text{or}}$/mK | $U$/mK |
| 1 | **K2**, Star | BNM-INM | 1985 (S) | BCMH2O | AL | 4N | -0,02 | 1,08 | -1,88 | 2,8 |
| 2 | **K2**, Int84, Star | CNR-IMGC | 1979 (S) | 3Ne | M-b (#2) | 4N5 (>4N5) | 0,11 | 0,28 | 0,11 | 0,28 |
| 3 | **K2** | KRISS | <1997 | open [g] | M | 4N4 | 0,01 | 0,4 | -0,15 | 0,4 |
| 4 | **K2** | NIST | 1979 | open | M (#8) | >5N5 | -0,13 | 0,32 | 0,04 | 0,32 |
| 5 | **K2**, Star, INRIM | NPL | 1993 (S) | Ne2 | AP (#18) | >5N | -0,10 | 0,44 | -0,19 | 0,38 |
| 6 | **K2, K2.1** | NRC | 1985 (S) | F15 | AP (#17) | 5N | -0,06 | 0,44 | -0,12 | 0,44 |
| 7 | **K2**, Star, INRIM | PTB | 1995 (S) | Ne-7 | L (#10) | 5N | — | — | 0,26 | 0,4 |
| 8 | **K2.1** | VNIIFTRI | — | open | — | — | 0,28 | 0,67 | 0,22 | 0,67 |
| 9 | **K2.3**, Star | INRIM (VSL) | 1999 (S) | 12Ne | MG (#11) | 5N | -0,645 | 0,66 | -0,71 | 0,66 |
| 10 | **K2.4**, Star | INM | 2002 (S) | Ne02/1 | AL (#14) | 5N (4N2) | -0,625 | 0,78 | -0,685 | 0,78 |
| 11 | **K2.4** | INRIM (INTiBS) | 2000 (S) | E3Ne | MG (#11) | 5N | -0,245 | 0,64 | -0,305 | 0,64 |
| 12 | **K2.3, K2.4, K2.5** | NRC | 2004 (S) | Cu-M-1 | P (#9) | 5N | -0,225 | 0,20 | -0,285 | 0,20 |
| 13 | **K2.3, K2.4, K2.5**, Star | NRC | 1985 (S) | F17 | AP (#17) | 5N | -0,133 | 0,20 | -0,173 | 0,20 |
| 14 | **K2.5**, INRIM | NMIJ-AIST | 2006 (S) | Ne-5 | AW (#7) | 5N (4N2) | -0,54 [f] | 0,60 | -0,60 [f] | 0,61 |
| 15 | **K2.5**, (K2.2), INRIM | INRIM | 2002 (S) | Ec2Ne | MG (#11) | 5N | -0,69 | 0,68 | -0,75 | 0,68 |
| 16 | (**K2.2**), Star | INRIM (NIM) | 2000 (S) | E2Ne | MG (#11) | 5N | K2.2 not yet completed | | | |
| 17 | Int84, Star, INRIM | CNR-IMGC | 1977 (S) | **1Ne** | M-a (#3) | 4N5 | | | | |
| 18 | Star | INRIM (PTB) | 1999 (S) | E1Ne | MG (#11) | 5N | | | | |
| 19 | Star | INRIM (DSIR) | 1986 (S) | 11Ne | S (#13) | 4N | | | | |
| 20 | Star | INRIM (INTiBS) | 2002 (S) | 7Ne | S (#13) | 4N | | | | |
| 21 | INRIM | NMIJ | 2005 (S) | Ne-2 | IB (#5) | 5N | | | | |
| 22 | INRIM | INRIM | 1999 (S) | 15Ne | MG (#11) | 5N | | | | |
| 23 | INRIM | INRIM | 2000 (S) | E4Ne | MG (#11) | 5N | | | | |
| 24 | INRIM | INRIM | 2001 (S) | Ec1Ne | MG (#11) | 5N | | | | |
| 25 | Int84, Star | VNIIFTRI | 1997 (S) | MC-897 | own | 5N | | | | |
| 26 | Int84 | NRC | 1979 (S) | Cell 12 | M | 4N5 | | | | |
| 27 | Int84 | INM | 1982 (S) | BCM4 | AL | 4N | | | | |
| 28 | Int84 | NRLM | 1978 (S) | 1Ne | J | 4N | | | | |
| 29 | Int84 | NRLM | 1978 (S) | 2Ne | J | 4N | | | | |

[a] K2 = CCT-K2 (NRC Pilot, 1997-99) [CCT-K2]; K2.1 to K2.5 = bi- and tri-lateral CCT-K2.x (NRC Pilot, 2003-2015) [CCT-K2.3; CCT-K2.4; CCT-K2.5]; Star = Comparison (PTB Co-ordinator, 2007-2012) [Fellmuth *et al.* 2012]; Int84 = International Intercomparison (IMGC-CNR Pilot, 1978-84) [Pavese *et al.* 1984]; INRIM = Comparison at INRIM (INRIM Co-ordinator, 2008-2009) [Pavese *et al.* 2010]. [b] Samples measured. [c] Assays reported in [Steur *et al.* 2017]. [d] Original differences with respect to the KCRV from [CCT-K2]. Set#: indicates thermometers set. [f] Mean of results. [g] Measurements repeated in 2013 with the isotopic composition known and corrected for it; the same thermometer has measured samples from two other bottles.

Table 2 – Data for neon used in this study and isotopic correction for CCT-K2.(x).

| # | Comparisons | Cell # | Set#1 K2.(x)[a] $T_{meas,bc}$/K $T_{KCRVbc}$ = 24,556 32 | Set#2 $T_{KCRVbc}$ = 24,556 38 | Corrections D$T$ (μK) Isotopic[d] | Chemical total[e] | Isotopic-corrected K2.(x)[b] $T_{meas,ac}$/K $T_{KCRVac}$ = 24,556 47 | $T_{KCRVac}$ = 24,556 56 | $T_{meas,ac}$− $T_{KCRVac}$ /μK | | Isotopic-corrected K2.(x)[c] $T_{thermal,ac}$/K $T_{th,mean}$ = 24,56610 K | $T_{th,mean}$ = 24,56610 K | ($T_{the}$ 24,5 / | |
|---|---|---|---|---|---|---|---|---|---|---|---|---|---|---|
| 1 | **K2**, Star | **BCMH2O** | 24,556 30 | 24,554 50 | 326 | 150 | 24,556 63 | 24,554 82 | 155 | -1737[g] | 24,555 97 | 24,554 17 | –166 | –2026[g] |
| 2 | **K2**, Int84, Star | **3Ne** | 24,556 43 | 24,556 49 | 310 | *47* | 24,556 74 | 24,556 80 | 269 | 237 | 24,556 12 | 24,556 18 | –20 | –20 |
| 3 | **K2** | **open**[f] | 24,556 33 | 24,556 23 | –34 | *1058* | 24,556 30 | 24,556 19 | –175 | –367 | 24,556 36 | 24,556 26 | 224 | 64 |
| 4 | **K2** | **open** | 24,556 19 | 24,556 42 | 22 | — | 24,556 21 | 24,556 44 | –259 | –121 | 24,556 17 | 24,556 39 | 28 | 198 |
| 5 | **K2**, Star, INRIM | **Ne2** | 24,556 22 | 24,556 19 | 279 | — | 24,556 50 | 24,556 46 | 28 | –94 | 24,555 94 | 24,555 91 | –199 | –289 |
| 6 | **K2, K2.1** | **F15** | 24,556 26 | 24,556 26 | 308 | — | 24,556 57 | 24,556 56 | 97 | 5 | 24,555 95 | 24,555 95 | –188 | –248 |
| 7 | **K2**, Star, INRIM | **Ne-7** | — | 24,556 64 | 196 | *17(1)*[h] | — | 24,556 83 | — | 273 | — | 24,556 44 | — | 244 |
| 8 | **K2.1** | **open** | 24,556 60 | 24,556 60 | 289 | — | — | 24,556 88 | — | 326 | — | 24,556 31 | — | 111 |
| 9 | **K2.3**, Star | **12Ne** | 24,555 68 | 24,555 67 | 123 | *24(1)*[h] | 24,555 80 | 24,555 79 | –673 | –770 | 24,555 55 | 24,555 54 | –588 | –653 |
| 10 | **K2.4**, Star | **Ne02/1** | 24,555 70 | 24,555 69 | –32 | *2(1)*[h] | 24,555 71 | 24,555 71 | –761 | –853 | 24,555 68 | 24,555 68 | –460 | –520 |
| 11 | **K2.4** | **E3Ne** | 24,556 08 | 24,556 07 | 123 | *24(1)*[h] | 24,556 20 | 24,556 19 | –273 | –365 | 24,555 95 | 24,555 95 | –188 | –248 |
| 12 | **K2.3, K2.4** | **Cu-M-1** | 24,556 10 | 24,556 09 | –32 | — | 24,556 09 | 24,556 08 | –382 | –474 | 24,556 10 | 24,556 10 | –39 | –99 |
| 13 | **K2.3, K2.4**, Star | **F17** | 24,556 21 | 24,556 20 | 308 | — | 24,556 52 | 24,556 51 | 44 | –48 | 24,555 90 | 24,555 89 | –241 | –301 |
| 14 | **K2.5**, INRIM | **Ne-5** | 24,555 78 | 24,555 78 | 4 | *6(5)*[h] | 24,555 78 | 24,555 78 | –207 | –299 | 24,555 78 | 24,555 78 | –364 | –419 |
| 15 | **K2.5,(K2.2),** INRIM | **Ec2Ne** | 24,555 63 | 24,555 63 | 123 | *24(1)*[h] | 24,555 75 | 24,555 75 | –357 | –449 | 24,555 51 | 24,555 51 | –633 | –688 |
| 17 | Int84, Star, INRIM | 1Ne | *24,556 43* | *24,556 49* | 115 | 272 | *24,556 55* | *24,556 60* | *74* | *42* | *24,556 28* | *24,554 07* | *–218* | *175* |
| 18 | Star | E1Ne | | | 123 | *24(1)*[h] | | | | | | | | |
| 19 | Star | 11Ne | | | 230 | 150 | | | | | | | | |
| 20 | Star | 7Ne | | | 230 | 150 | | | | | | | | |
| 21 | INRIM | Ne-2 | | | 23 | *24(1)*[h] | | | | | | | | |
| 22 | INRIM | 15Ne | | | 123 | *24(1)*[h] | | | | | | | | |
| 23 | INRIM | E4Ne | | | 123 | *24(1)*[h] | | | | | | | | |
| 24 | INRIM | Ec1Ne | | | 123 | *24(1)*[h] | | | | | | | | |
| 25 | Int84, Star | MC-897 | | | 289 | *343(1)*[h] | | | | | | | | |

Cell F15 is the reference cell for the CCT-K2.x. [a] Original, with KCRV$_{bc}$ computed as indicated in the text. [b] Using KCRV$_{ac}$ computed as indicated in the text (isotopic c only). [c] Isotopic-corrected values. [d] From [MeP 2014]. See uncertainties in column "$U_{iso,f}$" in Table 3. Mean for cells #2 to #8 is +180 μK. On isotopic assays, see also [S 2017]. [e] In italics the assays for only N$_2$ impurity (not available for H$_2$ impurity). No uncertainty: no reliable measure available. [f] See [g] in Table 1. [g] Not elaborated, clearl outlier. [h] If the impurity is higher than 10 ppm, the uncertainty of the impurity is 5 ppm or 2.5 % of the impurity itself, whichever is higher ($k$ = 1); if the impurity is low ppm, the uncertainty of the impurity is 50 % of the impurity itself ($k$ = 1).

Table 3. Uncertainties for neon of the inter-comparisons CCT-K2 and CCT-K2.x with and without the contribution of the isotopic factor.

| Comparisons | Cell fabricated (measured) | Sealing date | Cell | $U_{KC,or}$ [a] | | $U_{lab}$ mean [b] | | | $U_{CORR}$ mean [c] | | | better by [i] |
|---|---|---|---|---|---|---|---|---|---|---|---|---|
| | | | | #1 | #2 | $U_{iso}$ only [d] | $U$(Type B) | $U_{TOTlab}$ [e] | $U_{NET}$ without $U_{iso}$ [f] | $U_{iso,f}$ [g] | $U_{TRUE}$ [h] | |
| K2, Star | BNM-INM | 1985 | BCMH2O | 1080 [j] | 2800 | — | 1100 | 1100 | — | — | — | |
| K2, Int84, Star | CNR-IMGC | 1979 | 3Ne | 280 | 280 | 110 | 130 | 210 | 170 | 2 | 171 | 39% |
| K2 | KRISS | — | open | 400 | 400 | 300 [q] | 340 | 360 | 100 | 2 | 102 | 74% |
| K2 | NIST | — | open | 320 | 320 | 150 | 180 | 265 | 170 | 2 | 171 | 46% |
| K2, Star, INRIM | NPL | 1993 | Ne2 | 440 | 380 | 300 | 190 | 200 | 80 | | 80 | 79% |
| K2, K2.1 | NRC | 1985 | F15 | 440 | 440 | 320 [q] | 360 | 400 | 120 | | 120 | 73% |
| K2, Star, INRIM | PTB | 1995 | Ne-7 | | 400 | 320 | 340 | 360 | 80 | 2 | 108 | 73% |
| K2.1 | VNIIFTRI | — | open | 670 | | 400 [q] | — | 420 | 270 | | | |
| K2.3, Star | INRIM (VSL) | 1999 | 12Ne | 620 | 620 | 314 | — | 392 | 306 | 8 | 310 | 50% |
| K2.4, Star | INM | 2002 | Ne02/1 | 760 | 760 | 480 | 600 | 600 | 280 | 2 | 289 | 62% |
| K2.4 | INRIM (INTiBS) | 2000 | E3Ne | 640 | 640 | 320 [q] | — | 400 | 320 | 8 | 324 | 49% |
| K2.3, K2.4 | NRC | 2004 | Cu-M-1 | 640 | 640 | *320* | — | 400 | 320 | 4 | 334 | 48% |
| K2.3, K2.4, Star | NRC | 1985 | F17 | 640 | 640 | *320* | — | — | 320 | | 320 | 50% |
| K2.5, INRIM | NMIJ-AIST | 2006 | Ne-5 | 600 | 600 | 280 [r] | — | 290 | 320 | | 320 | 50% |
| K2.5, (K2.2), INRIM | INRIM | 2002 | Ec2Ne | 680 | 680 [s] | 8 [t] | — | 110 | — | 6 | — | — |
| s.d. [k] | | | | | | | | | | | | 15% [p] |
| Mean [m] | | | | | | | | | | | | 60% [p] |

[a] Original uncertainties of the $\Delta T$s ($\Delta T_{or}$ in Table 1). [b] Original laboratory uncertainty budgets from [CCT-K2]: mean of #1 and #2 sets. [c] Uncertainties [a] cleaned up from isotopic uncertainties ("$U_{iso}$ only" under [b]): mean on #1 and #2 sets. [d] Item for the isotopic effect of the *estimated* laboratory uncertainty. [e] Total laboratory uncertainty. [f] Uncertainty budget corrected for the isotopic effect contribution. [g] Uncertainty of the isotopic corrections taken from [Steur *et al.* 2017]. [h] New total uncertainty budget into account. [i] Lowering of [h] with respect to [e]. [j] Grayed data not elaborated. [k] Standard deviation of the column data. [m] Mean of the column data. [p] Improvement of [h] o
[q] Also includes chemical impurities. [r] Uncorrected, used in the uncertainty budget. [s] For the direct comparison INRIM-NRC 300 μK. [t] Uncertainty of the corrected val

Table 4. Data for neon used in this study: isotopic correction for the Star cell intercomparison.

| Comparisons | # | Cell, fabricated (measured) | Sealing date | Cell | Gas (# analysis) | STAR (Ref: PTB Ne-7) | |
|---|---|---|---|---|---|---|---|
| | | | | | | D$T$ = cell-ref | D$T$ corrected |
| K2, **Star** | 1 | BNM-INM | 1985 | **BCMH2O** | AL1 (#19) | 210 | 80 |
| Int84, **Star** | 3 | CNR-IMGC | 1979 | **3Ne** | M-b (#2) | 66 | -49 |
| K2 | 4 | (KRISS) | open | open | M | | |
| K2 | 5 | (NIST) | open | open | M (#8) | | |
| K2, **Star**, INRIM | 6 | NPL | 1993 | **Ne2** | AP (#18) | 132 | 49 |
| K2, K2.1 | 7 | NRC | 1985 | F15 | AP (#17) | | |
| K2, **Star**, INRIM | 8 | **PTB** | 1995 | **Ne-7** [a] | L (#10) | 0 [a] | **0** |
| K2.3, **Star** | 10 | INRIM (VSL) | 1999 | **12Ne** | MG (#11) | -62 | 9 |
| K2.4, **Star** | 11 | INM | 2002 | **Ne02/1** | AL2 (#14) | 8 | 236 |
| K2.4 | 12 | INRIM (INTiBS) | 2000 | E3Ne | MG (#11) | | |
| K2.3, K2.4 | 13 | NRC | 2004 | Cu-M-1 | P (#9) | | |
| K2.3, K2.4, **Star** | 14 | NRC | 1985 | **F17** | AP (#17) | 170 | 58 |
| (K2.2) | 15 | INRIM (NIM) | 2000 | E2Ne | MG (#11) | | |
| K2.5, INRIM | 16 | NMIJ | 2006 | Ne-5 | AW (#7) | | |
| INRIM | 17 | NMIJ | 2005 | Ne-2 | IB (#5) | | |
| K2.5, (K2.2), INRIM | 18 | INRIM | 2002 | Ec2Ne | MG (#11) | | |
| Int84, **Star**, INRIM | 2 | CNR-IMGC | 1977 | **1Ne** | M-a (#3) | 37 | 117 |
| **Star** | 19 | INRIM (PTB) | 1999 | **E1Ne** | MG (#11) | -23 | 48 |
| **Star** | 20 | INRIM | 1986 | **11Ne** | S (#13) | 55 | 21 |
| INRIM | 21 | INRIM | 1999 | 15Ne | MG (#11) | | |
| INRIM | 22 | INRIM | 2000 | E4Ne | MG (#11) | | |
| INRIM | 23 | INRIM | 2001 | Ec1Ne | MG (#11) | | |
| **Star** | 24 | INRIM (INTiBS) | 2002 | **7Ne** | S (#13) | 77 | 43 |
| Int84, **Star** | 25 | VNIIFTRI | 1997 | **MC-897** | own | 156 | 63 |
| Int84 | 26 | NRC | 1979 | Cell 12 | M | | |
| Int84 | 27 | INM | 1982 | BCM4 | AL | | |
| Int84 | 28 | NRLM | 1978 | 1Ne | J | | |
| Int84 | 29 | NRLM | 1978 | 2Ne | J | | |
| **Star** | 30 | INM | 1999 | **Ne99/2** | AL2 (#19) | 205 | 75 |
| **Star** | 31 | PTB | 1995 | **Ne-12** | L (#10) | 8 | 8 |
| **Star** | 32 | NIST | 1998 | **NIST201** | Math (#20) [b] | 130 | *302* [b] |

[a] Reference cell. [b] Uncertain filling-gas attribution to a bottle.